\documentclass{PoS}
\usepackage{subfigure}
\usepackage{cite}
\newcommand{\beq}{\begin{equation}}
\newcommand{\eeq}{\end{equation}}
\newcommand{\bea}{\begin{eqnarray}}
\newcommand{\eea}{\end{eqnarray}}

\renewcommand{\d}{\delta}
\renewcommand{\l}{\lambda}

\renewcommand{\b}{\beta}

\renewcommand{\k}{\kappa}

\newcommand{\s}{\sigma}

\newcommand{\oh}{{\textstyle{\frac{1}{2}}}}

\newcommand{\oq}{{\textstyle{\frac{1}{4}}}}

\newcommand{\non}{\nonumber}

\newcommand{\rf}[1]{(\ref{#1})}
\newcommand{\ra}{\rightarrow}
\newcommand{\pa}{\partial}

\title{Yang-Mills Ground State in 2+1 Dimensions and Temporal Gauge}

\ShortTitle{Yang-Mills Ground State in 2+1 Dimensions}

\author{\speaker{Jeff Greensite}\thanks{Research supported in part by 
the U.S.\ Department of Energy under Grant No.\ DE-FG03-92ER40711.} \\
       Physics and Astronomy Dept.,
       San Francisco State University,
       San Francisco, CA 94132 USA \\
       E-mail: \email{greensit@stars.sfsu.edu}}

\author{{\v S}tefan Olejn\'{\i}k\thanks{Research supported in part by 
the Slovak Science and Technology Assistance Agency under Contract 
No.\ APVT--51--005704, and the Grant Agency for Science, Project 
VEGA No.\ 2/6068/2006.}  \\
Institute of Physics, Slovak Academy
of Sciences, SK--845 11 Bratislava, Slovakia \\
E-mail: \email{stefan.olejnik@savba.sk}}

\abstract{A gauge-invariant wavefunctional is proposed as
an approximation to the ground state of Yang-Mills theory 
in 2+1 dimensions, quantized in temporal gauge.  The proposed
vacuum state is the true ground state of the appropriate Hamiltonian 
in both the free-field limit, and in a zero mode strong-field limit.  
Confinement, in this approach, arises via dimensional reduction, and 
we present numerical results for the mass gap.  The issue of color 
screening is briefly discussed.}

\FullConference{The XXV International Symposium on Lattice Field Theory\\
                July 30 - August 4 2007\\
                Regensburg, Germany}

\begin{document}

\section{Introduction}

From time to time there have been efforts \cite{Me1,Me2,Marty,
Feynman,Mansfield,Kovner,Samuel,Adam,Hugo,Peter,KKN,Rob,
Cornwall} 
to obtain the Yang-Mills vacuum wavefunctional, often in lower 
dimensions, to see if anything can be learned about confinement 
and the mass spectrum.  The problem is to solve for the ground state 
$\Psi_0[A]$ of the Yang-Mills Hamiltonian $H$, which has its simplest 
form in temporal gauge   
\beq
H = \int d^dx \left\{ -\oh {\d^2 \over \d A^a_k(x)^2} + \oq 
F_{ij}^{a}(x)^2 \right\}
\eeq
The price for this simplicity, in temporal gauge, is that all physical 
states are required to satisfy a constraint
\beq
\Bigl( \d^{ac} \pa_k + g \epsilon^{abc} A^b_k \Bigr) {\d \over \d A^c_k }\Psi = 0 
\eeq
which enforces invariance under infinitesimal gauge transformations.

Our claim is that the ground state solution in $D=2+1$ dimensions, in temporal 
gauge, is approximated by\footnote{Closely related proposals have been made in 
the past by Samuel \cite{Samuel} and Diakonov \cite{Mitya}, cf.\ the discussion 
in ref.\ \cite{Us}.}  
\beq
\Psi_0[A] = \exp\left[-\oh\int d^2x d^2y ~ B^a(x) \left({1 \over 
\sqrt{-D^2 -  \l_0   + m^2}} \right)^{ab}_{xy} B^b(y) \right]
\label{vac}
\eeq
where $B^a=F_{12}^a$ is the color magnetic field strength,
$D^2$ is the covariant Laplacian in the adjoint color representation,
$\l_0$ is the lowest eigenvalue of $-D^2$, and $m$ is a constant of
order $g^2$. In support of this claim, we will argue that \rf{vac}
(i) is a solution of the YM Schr{\"o}dinger equation in the $g\ra 0$ limit;
(ii) solves the YM Schr{\"o}dinger equation in the strong field, zero-mode
 limit; (iii) confines if  $m > 0$, and that $m > 0$ seems energetically preferred; 
 and (iv) results in the numerically correct relationship between the mass 
 gap and the string tension.  A more complete report of our results can be 
 found in a recent article \cite{Us}.
 
\section{Free-Field and Zero-Mode Limits}

    In the free-field $g=0$ limit, eq.\ \rf{vac} becomes
\beq
\Psi_0[A] = \exp\left[-\oh\int d^2x d^2y \Bigl(\pa_1 A^a_2(x) - \pa_2 A^a_1(x) \Bigr) 
\left({\d^{ab} \over \sqrt{-\nabla^2}} \right)_{xy} 
\Bigl(\pa_1 A^b_2(y) - \pa_2 A^b_1(y) \Bigr) \right]
\eeq
which is the known ground state solution in the abelian, free-field case, and 
an obvious starting point for any investigation such as ours.  Apart from the 
free-field limit, there is another, quite different limit which can be treated 
analytically.   Consider gauge fields which are constant in space, but variable 
in time, in D=2+1 dimensions.   The Lagrangian is
\bea
      L &=& \oh \int d^{2}x ~ \Bigl[ \pa_{t} A_{k}\cdot \pa_{t} A_{k}
             - g^{2} (A_{1} \times A_{2}) \cdot (A_{1} \times A_{2}) \Bigr]
\non \\
      &=&  \oh V \Bigl[ \pa_{t} A_{k}\cdot \pa_{t} A_{k}
             - g^{2} (A_{1} \times A_{2}) \cdot (A_{1} \times A_{2}) \Bigr]
\eea
with Hamiltonian operator
\beq
     H = -\oh {1\over V} {\pa^{2} \over \pa A_{k}^{a} \pa A_{k}^{a} }
                 + \oh g^{2} V  (A_{1} \times A_{2}) \cdot (A_{1} \times A_{2}) 
\eeq
where $V$ is the volume of 2-space.
With some algebra, one may verify that
\beq
           \Psi_{0} = \exp\left[- \oh g V {(A_{1} \times A_{2}) \cdot 
                      (A_{1} \times A_{2}) \over \sqrt{|A_{1}|^{2} + |A_{2}|^{2}}}  \right]
\label{zm}
\eeq
satisfies the zero-mode Yang-Mills Schr{\"o}dinger equation up to
$1/V$ corrections.   For comparison, we consider our proposed
vacuum state \rf{vac} in the strong $A$-field limit, where the covariant
Laplacian is dominated by the gauge-field zero-mode, i.e.  
\beq
          (-D^{2})^{ab}_{xy} =
 g^{2} \d(x-y) \Bigl[  (A_{1}^{2}+A_{2}^{2})\d^{ab} - A_{1}^{a}A_{1}^{b} -    A_{2}^{a}A_{2}^{b} \Bigr]   
\eeq
In this limit, it is not hard to show that $\Psi_0[A]$ of eq.\ \rf{vac} precisely 
reduces to \rf{zm}.  Thus the proposed ground state not only has the correct 
free-field limit, but also agrees with the calculable ground state of the zero-mode 
Schr{\"o}dinger equation in an appropriate strong-field limit.

\section{Dimensional Reduction and Confinement}

A long time ago it was suggested \cite{Me1} that at large distance scales, the pure 
Yang-Mills vacuum in a confining theory looks like
\beq
\Psi_0^{eff} \approx \exp\left[ -\mu \int d^dx ~ F^a_{ij}(x) F^a_{ij}(x) \right]
\eeq
This vacuum state has the property of \emph{dimensional reduction}:  Computation 
of a spacelike loop in $d+1$ dimensions reduces to the calculation of a 
Wilson loop in Yang-Mills theory in $d$ Euclidean dimensions.  In $d=2$ 
dimensions the Wilson loop can be calculated analytically, and we know 
there is an area-law falloff, with Casimir scaling of the string tensions.  Now 
suppose, in the proposed vacuum \rf{vac}, we expand the $B$-field in 
eigenmodes of the covariant Laplacian, i.e.\ $B^a(x) = \sum_n b_n \phi_n^a(x)$ 
where $-D^2  \phi_n^a(x) = \l_n  \phi_n^a(x)$.  Define the ``slow" component
\beq
B^{a,{\rm slow}}(x) = \sum_{n=0}^{n_{max}} b_{n} \phi^{a}_{n}(x)
\eeq
with mode cutoff $n_{max}$ defined such that $\l_{n_{max}}-\l_0 \ll m^2$.  Then 
the portion of the squared wavefunctional \rf{vac} which is quadratic in $B^{slow}$, 
to leading order in $1/m^2$, is
\beq
|\Psi_0|^2 = \exp\left[-{1\over m}\int d^2x ~ B^{slow} B^{slow}\right]
\eeq
which has the dimensional reduction form.   The string tension for fundamental 
representation Wilson loops is easily computed in two Euclidean dimensions, and 
in lattice units it is $\s=3m/(4\b)$.  Suppose we turn this around, and choose the 
mass parameter $m={4\over 3} \b \s$.  Then our proposed vacuum wavefunctional 
must imply a definite mass gap, which we would like to calculate.

\section{The Mass Gap}

To get the mass gap, we need to compute the connected correlator
\beq
   \langle B^2(x) B^2(y) \rangle_{conn} \equiv  
   \langle (B^a B^a)_x (B^b B^b)_y \rangle - \langle (B^a B^a)_x \rangle^2
\eeq
in the probability distribution
\beq
P[A] = |\Psi_0[A]|^2 = \exp\left[-\int d^2x d^2y ~ B^a(x) K^{ab}_{xy}[A] B^b(y) \right]
\eeq
where
\beq
K_{xy}^{ab}[A] = \left({1 \over \sqrt{-D^2 -   \l_0  + m^2}} \right)^{ab}_{xy}  
\eeq
Numerically this looks hopeless!  Not only is the kernel $K_{xy}^{ab}$           
highly non-local, but it is not even known explicitly for arbitrary gauge fields.   
We have, nonetheless, found a way of carrying out the numerical simulation, 
which relies on the fact that, after gauge-fixing, the variance in the kernel among 
thermalized configurations is negligible.  The method is described in ref.\ \cite{Us};
 here we will only present the results.

Observables of interest include the eigenvalue spectrum $\{\l_n\}$ of the adjoint 
covariant Laplacian, and the connected field-strength correlator 
$\langle B^2(x) B^2(y) \rangle_{conn} \propto G(x-y)$, where
\beq
G(x-y) = \Bigl\langle (K^{-1})^{ab}_{xy}(K^{-1})^{ba}_{yx} \Bigr\rangle
\eeq
Choosing $m={4\over 3} \b \s$, the mass gap can be extracted from
$G(R)$.  For comparison, we can also compute these observables on
time-slices of lattices generated by ordinary lattice Monte Carlo in $D=3$ 
Euclidean dimensions.  We will refer to these time-slices as \emph{MC lattices}, 
and they can be regarded as drawn from the probability distribution $|\Psi_0^E[U]|^2$,
where $\Psi_0^E$ is the ground state of the transfer matrix of the Wilson action.  
Lattices generated by the method described in ref.\ \cite{Us}, which simulates 
our proposed ground state wavefunctional, will be referred to as \emph{recursion lattices}.

\begin{figure}
\centering
\subfigure[] 
{
    \label{fig1:eig}
    \includegraphics[width=6.5cm]{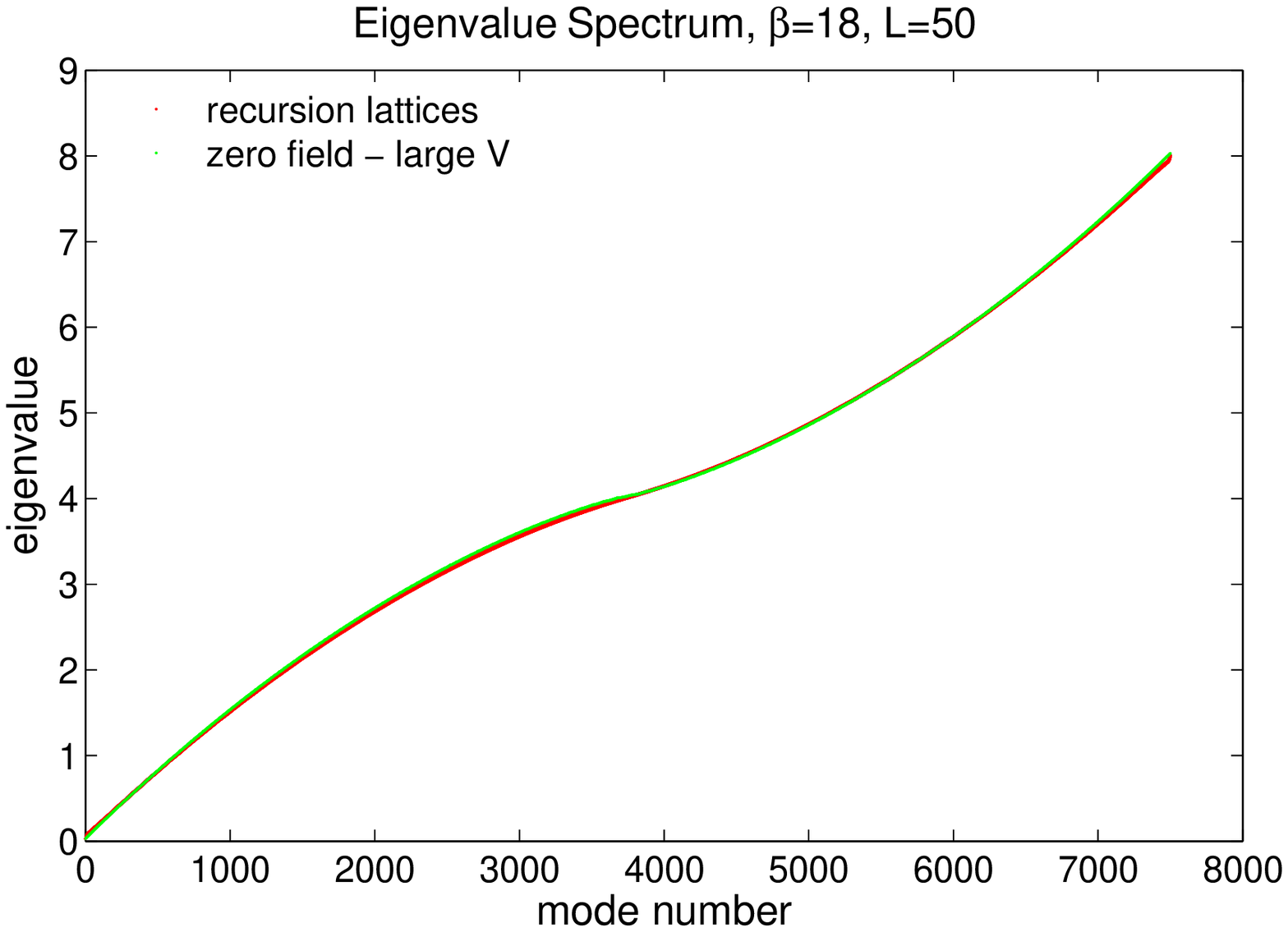}
}
\hspace{0.5cm}
\subfigure[] 
{
    \label{fig1:mass}
    \includegraphics[width=7.2cm]{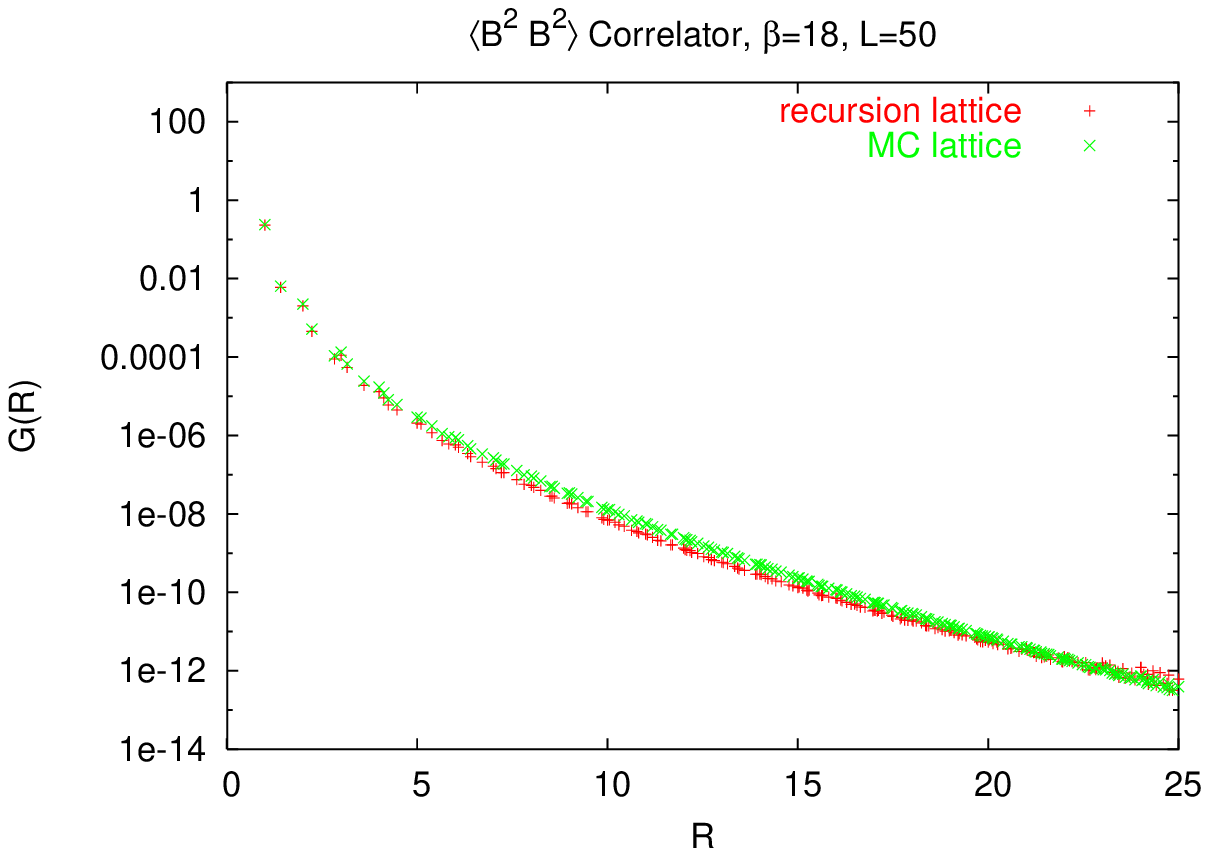}
}
\caption{Simulations of the vacuum state at $\b=18$ and lattice extension $L=50$.  
(a) Eigenvalue spectra of the operator $-D^{2} -\l_{0}+m^{2}$.  Also plotted is the 
spectrum of the large-volume zero-field operator $-\nabla^{2} + m^{2}$.  (b) The 
correlator $G(R)$ computed on recursion and MC lattices.}
\label{fig1} 
\end{figure}

\begin{figure}
\centering
\subfigure[] 
{
    \label{fit}
    \includegraphics[width=6.8cm]{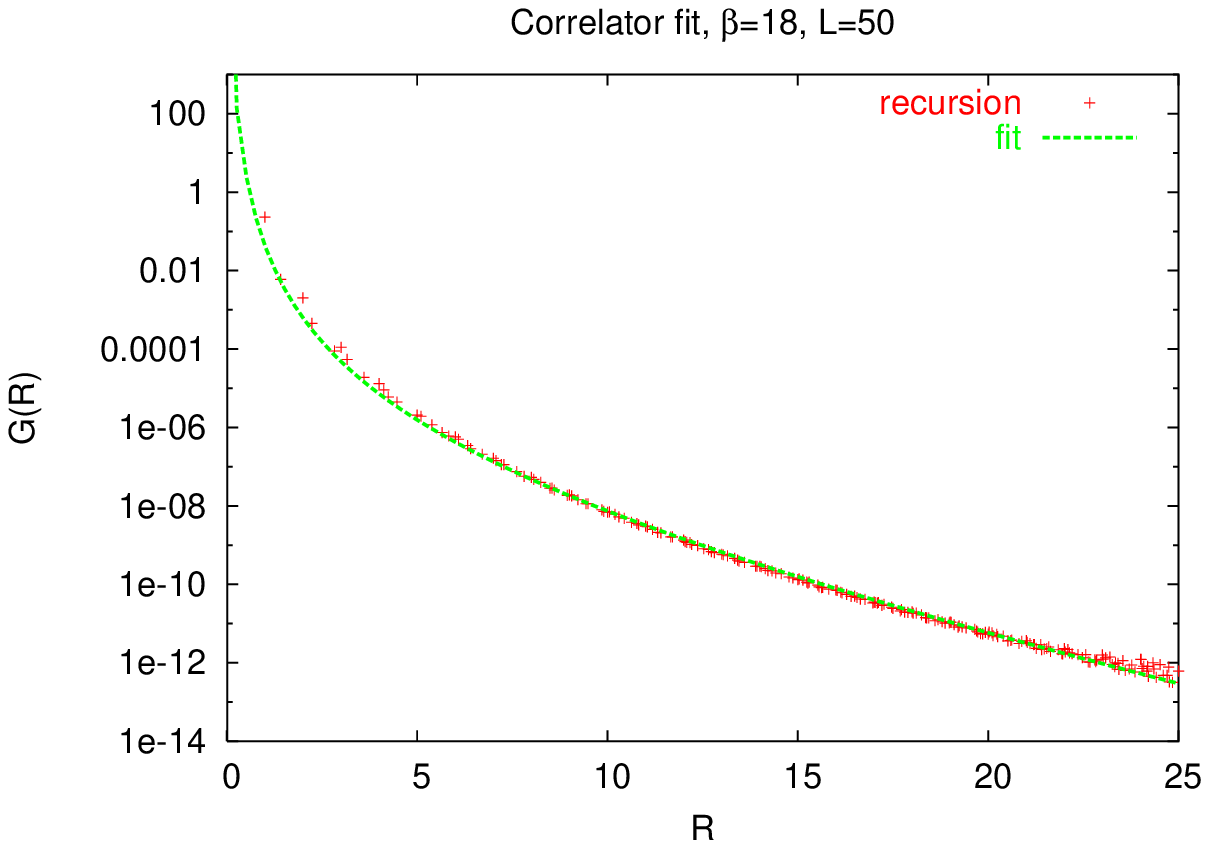}
}
\hspace{0.5cm}
\subfigure[] 
{
    \label{gap}
    \includegraphics[width=6.8cm]{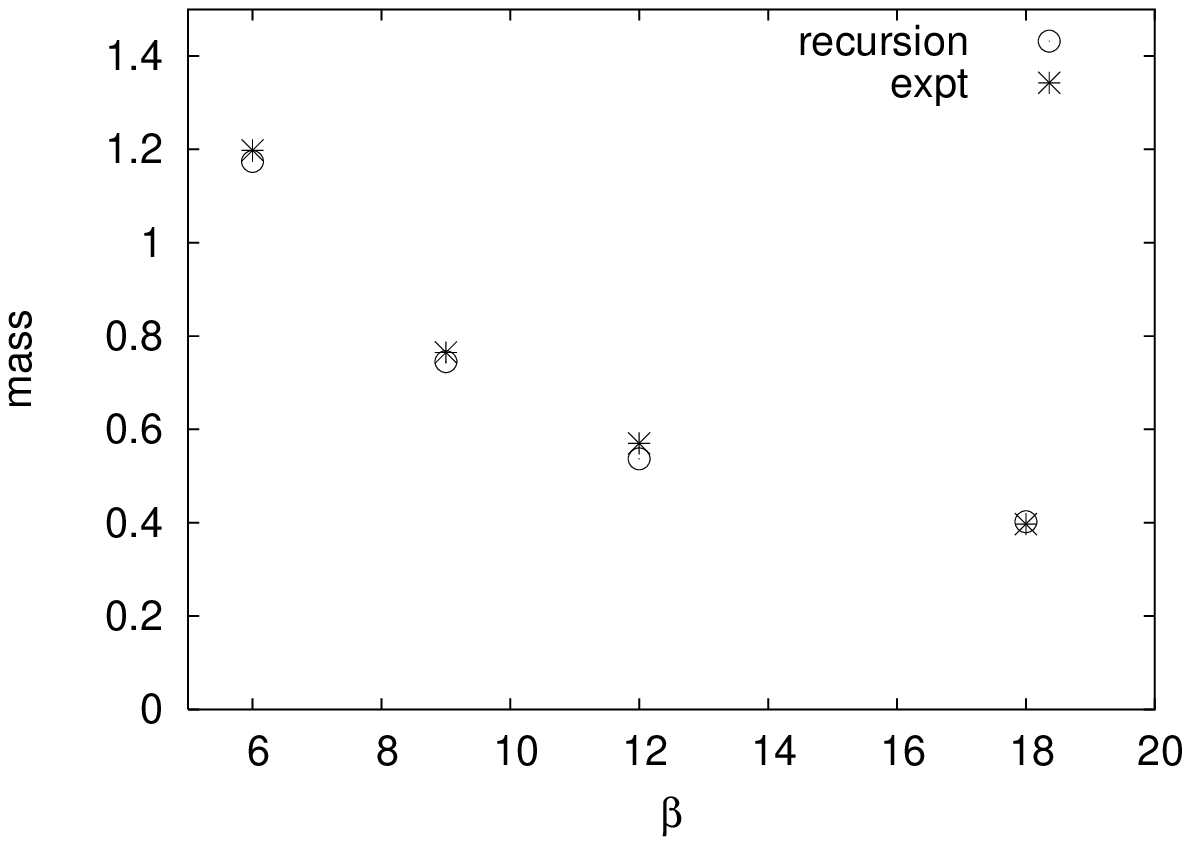}
}
\caption{(a) Best fit (dashed line) of the recursion lattice data for 
$G(R)$ by the analytic form given in eq.\  (4.5).  
(b) Mass gaps extracted from recursion lattices at various lattice 
couplings, compared to the $0^+$ glueball masses in 2+1 
dimensions obtained in ref.\  \cite{Teper} (denoted ``expt") 
via standard lattice Monte Carlo methods. Errorbars are 
smaller than the symbol sizes.
}
\label{fig2} 
\end{figure}

    Fig.\ \ref{fig1:eig} is a plot of eigenvalue vs.\ mode number for the operator 
$-D^{2} -\l_{0}+m^{2}$, at $\b=18$, from ten independent $50\times 50$ 
recursion lattices (no averaging).   Also plotted, but indistinguishable from 
the other spectra, is the rescaled spectrum of the large-volume zero-field 
operator $-\nabla^{2} + m^{2}$.  It is obvious from this figure that there is 
very little variance in the spectrum of $-D^2 -\l_0$ from one thermalized 
lattice to the next, moreover, this spectrum is very close to that of a free-field 
theory.  Fig.\ \ref{fig1:mass} shows our data for $G(R)$, obtained from ten 
recursion lattices, and ten MC lattices.  Note the tiny values of 
$G(R) \sim 10^{-12}$ obtained at the larger R values. This 
requires a near-absence of fluctuation in $K^{-1}$ from one 
thermalized lattice to the next.  Note also that $G(R)$ obtained on MC and 
recursion lattices agree very closely with one another.  The mass gap $M$ 
is extracted from a two-parameter $(c,M)$ fit of the recursion lattice data to the 
function\footnote{This expression is motivated from the functional 
form of $K_{0xy}^{-1}K_{0yx}^{-1}$ where $K_{0}^{-1} = \sqrt{-\nabla^2 +m^2}.$ }
\beq
G_{fit}(R) = c (1 + \oh M R)^2 {e^{-MR} \over R^6}
\label{G0}
\eeq
The best fit to the data for $G(R)$ at $\b=18$ is shown in Fig.\ \ref{fit}.  
Our final results for the mass gaps at a variety 
of $\b$ values are shown in Fig.\ \ref{gap}, together with the results for the $0^{+}$ 
glueball obtained by Meyer and Teper \cite{Teper}, using standard methods.  The 
agreement is clearly very good.

\section{Why confinement?}

    We have seen that there is a mass gap if $K^{-1}$ has a finite range, 
and a non-zero string tension (by the dimensional reduction argument) 
if $K$ is also finite range.  Both of these conditions are obtained if the 
mass parameter $m$ is non-zero, so the question "why confinement?" 
boils down to the question: why is $m$ non-zero? An obvious approach 
is to treat $m$ as a variational parameter, and use it to minimize 
$\langle H\rangle$ .    To simplify matters, noting the negligible 
fluctuation in $K$, we make the drastic approximations of (i) neglecting 
functional derivatives of the kernel  $K$ in  computing $\langle H\rangle$; and 
(ii) ignoring correlations between $B$ and $K$.  With these approximations, 
the VEV of $H$ turns out to be
\beq
\langle H \rangle   =  {1\over 2} \left\langle \mbox{Tr} \sqrt{-D^2 - \lambda_{0} + m^2}    
+ {1\over 2} \mbox{Tr}{ \lambda_{0}-m^{2} \over \sqrt{-D^2 - \lambda_{0} + m^2}} \right\rangle  
\eeq
There is a competition between the two terms on the rhs; the first
increases with $m^2$, while the latter decreases.  In an abelian theory, where 
$D^2=\nabla^2$ and $\l_0=0$, it is trivial to show that $\langle H\rangle$ 
is minimized at $m=0$, implying no confinement, and no mass gap in the 
abelian case. For a non-abelian theory matters are different, chiefly because 
$\l_0>0$. A naive calculation, which ignores any dependence of the $\{\l_n\}$ 
on $m^2$, finds that $m^2 = \langle \l_0 \rangle$ at the minimum.  While a more 
accurate calculation would no doubt change somewhat the numerical value of 
$m^2$ at the minimum, it seems unlikely that this value would be shifted exactly 
to zero. 

\section{What about N-ality?}

    Dimensional reduction at large scales, from three to two dimensions, implies 
Casimir scaling of the string tensions; in particular, the adjoint string tension is
non-zero.  While this is correct at intermediate 
distance scales, asymptotically the string tension should depend only on the 
N-ality of the representation.  The transition from Casimir scaling to N-ality 
dependence must be coming from corrections to dimensional reduction.
    
\begin{figure}
\centering
\subfigure[] 
{
    \label{r}
    \includegraphics[width=8.5cm]{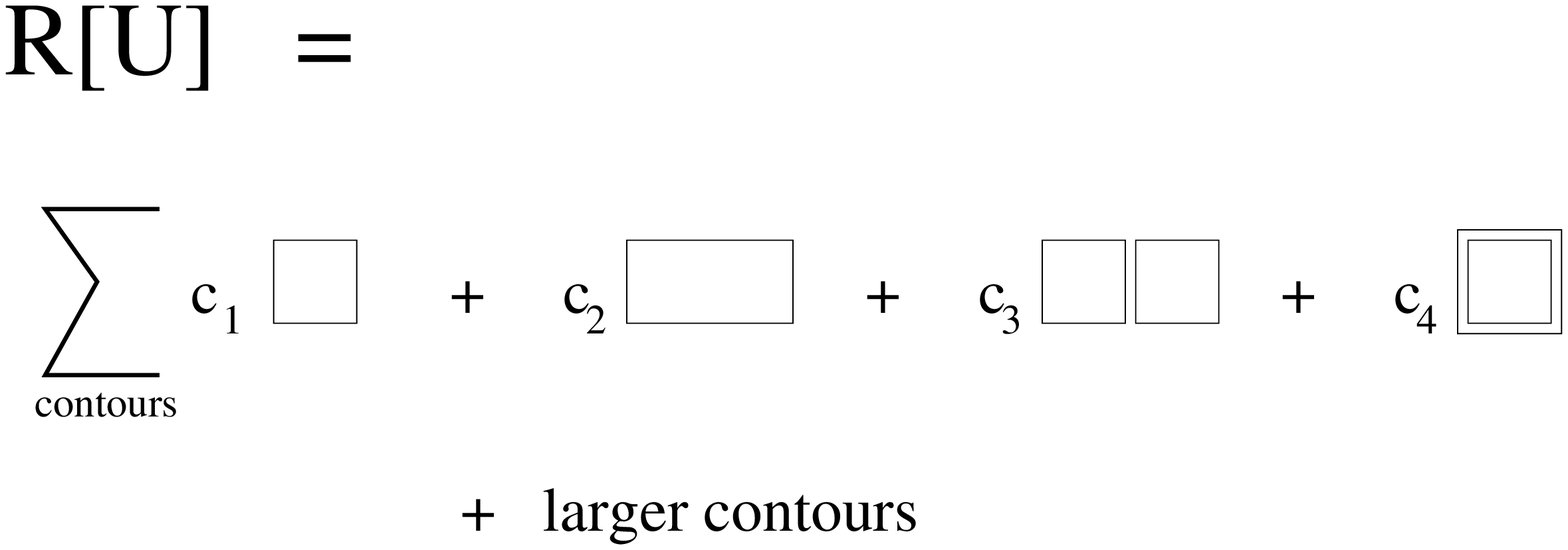}
}
\hspace{0.5cm}
\subfigure[] 
{
    \label{tile}
    \rotatebox{90}{\includegraphics[width=4cm]{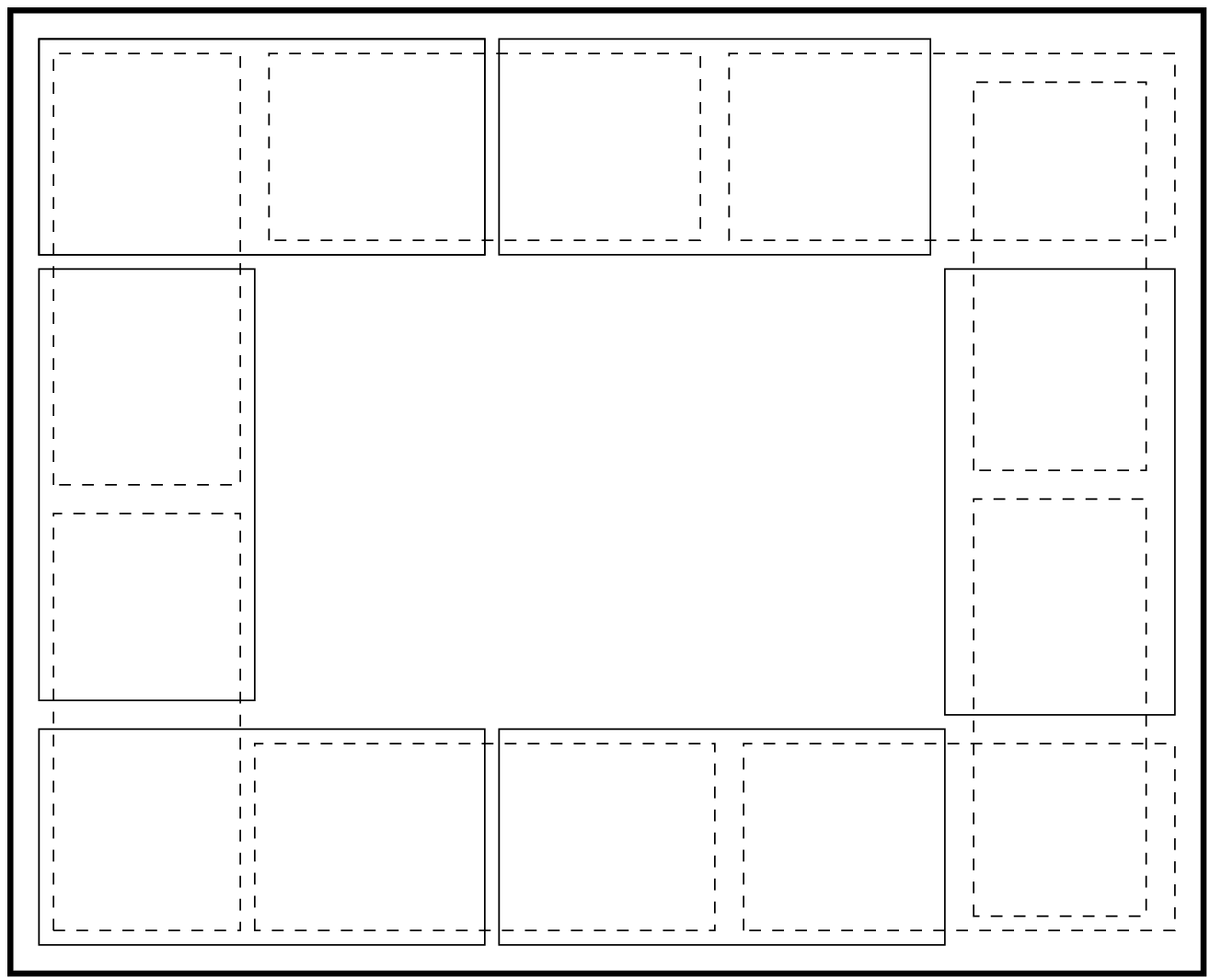}}
}
\caption{(a) The first few terms in the strong-coupling expansion of 
the lattice vacuum state $\Psi_0[U]$, with $R[U]=\log(\Psi_0[U])$. 
(b) How $1\times 2$ rectangles in $R[U]$ screen an adjoint
Wilson loop.  The adjoint Wilson loop is denoted by a heavy solid line. 
The overlapping $1\times 2$ rectangles are indicated 
by (alternately) light solid and light dashed lines.}
\label{fig:sub} 
\end{figure}

    It is worth examining how N-ality dependence is achieved when the vacuum state is 
computed by the strong-coupling method developed in ref.\ \cite{Me2}.  In this 
case the ground state has the form $\Psi_0[U]=\exp[R[U]]$, where $R[U]$ is a sum 
over contours (plaquette, $1\times 2$ rectangle, etc.) constructed from $p$ plaquettes, 
with each contour multiplied by a constant of order $\b^{2p}$.  The leading contributions, 
at strong-coupling, are shown in fig.\ \ref{r}.  It is not hard to show that the perimeter 
law for large adjoint loops is obtained, in the strong-coupling expansion of $\Psi_0^2[U]$, 
by tiling the perimeter of the loop with overlapping $1\times 2$ rectangles, as shown 
in Fig.\ \ref{tile}.

It is interesting that the rectangle term not only screens adjoint loops, but 
is also responsible for the leading correction to dimensional reduction.  An 
expansion of the strong-coupling vacuum in powers of lattice spacing 
yields \cite{Guo}
\beq
\Psi_0[U] =
 \exp \left[ - {2\over \b} \int d^2x ~( a \k_0 B^2 - a^3 \k_2 B(-D^2)B + \ldots ) \right] 
\eeq
The first term represents dimensional reduction, and all four contours shown 
in Fig.\ \ref{r} contribute to $\k_0$.  Only the rectangle contour, however, 
contributes to the second term, which supplies the leading correction to 
dimensional reduction.  It is also the rectangle contour which couples together 
field strengths in neighboring plaquettes.

Returning to our proposed vacuum state and expanding the kernel
in powers of $1/m^2$, the part of $\Psi_0^2$ quadratic in
$B^{slow}$, up to next-to-leading order in $1/m^2$, is
\beq
     \exp\left[-{1\over m}\int d^2x ~ \left( B^{slow} B^{slow}
             - B^{slow}{-D^2 -\l_0\over 2m^2}B^{slow} + \ldots \right) \right]
\eeq
Note that the correction to the first (dimensional reduction) term is similar to, 
and has the same sign as, the leading correction to dimensional reduction in 
the strong-coupling vacuum state, generated  by the rectangle term.  Our 
speculation is that, as in strong coupling, this second term is associated with 
the screening of higher-representation color charges, converting Casimir scaling 
to N-ality dependence of the higher-representation string tensions.

\end{document}